\documentclass[default,round]{sn-jnl}
\usepackage{graphicx}
\usepackage{url}
\usepackage{natbib}

\usepackage{booktabs} % put this in your preamble
\usepackage{url}
\usepackage{doi}
\usepackage{graphicx}%

\usepackage{float} 
\usepackage{multirow}%
\usepackage{placeins}
\usepackage{amsmath,amssymb,amsfonts}%
\usepackage{amsthm}%
\usepackage{mathrsfs}%
\usepackage[title]{appendix}%
\usepackage{xcolor}%
\usepackage{textcomp}%
\usepackage{manyfoot}%
\usepackage{booktabs}%
\usepackage{algorithm}%
\usepackage{algorithmicx}%
\usepackage{algpseudocode}%
\usepackage{listings}%
\usepackage[draft]{changes}
\usepackage{enumitem}

\usepackage{hyperref}

\usepackage{titlesec}

\titleformat{\subparagraph}[runin]{\bfseries}{\thesubparagraph}{1em}{}

\usepackage{todonotes}
\usepackage{comment}
\usepackage{color, colortbl}	

\usepackage{adjustbox}
\NewDocumentCommand{\rot}{O{90} O{1em} m}{\makebox[#2][l]{\rotatebox{ #1}{#3}}}%

\definecolor{mygray}{gray}{0.9}
\newcolumntype{g}{>{\columncolor{mygray}}c}

\usepackage{changes}
\usepackage{xcolor}
\usepackage{csquotes}
\definechangesauthor[name=Gerhard, color=magenta]{GH}
\definechangesauthor[name=Esra, color=brown]{EY}
\definechangesauthor[name=All, color=orange]{ALL}

\usepackage{ulem}

\theoremstyle{thmstyleone}%
\theoremstyle{thmstyletwo}%

\theoremstyle{thmstylethree}%

\begin{document}

\title{Quantum Leap in Finance: Economic Advantages, Security, and Post-Quantum Readiness}
%\title{Quantum Computing in Finance: How to identify an economic advantage}
\author[1]{\fnm{Gerhard} \sur{Hellstern}}\email{gerhard.hellstern@dhbw-stuttgart.de}
\author[2]{\fnm{Esra} \sur{Yeniaras}}\email{esye@ek.dk}

\equalcont{All authors contributed equally to this work.}

\affil[1]{\orgdiv{Center of Finance}, \orgname{Cooperative State University of Baden-Württemberg (DHBW)}, \orgaddress{\city{Stuttgart}, \country{Germany}}}

\affil[2]{\orgdiv{Cyber Security Department}, \orgname{Copenhagen Business Academy (EK)}, Copenhagen, \orgaddress{\country{Denmark}}}

\abstract{This paper provides an in-depth review of the evolving role of quantum computing in the financial sector, emphasizing both its computational potential and cybersecurity implications. Distinguishing itself from existing surveys, this work integrates classical quantum computing applications—such as portfolio optimization, risk analysis, derivative pricing, and Monte Carlo simulations—with a thorough examination of blockchain technologies and post-quantum cryptography (PQC), which are crucial for maintaining secure financial operations in the emerging quantum era. We propose a structured four-step framework to assess the feasibility and expected benefits of implementing quantum solutions in finance, considering factors such as computational scalability, error tolerance, data complexity, and practical implementability. This framework is applied to a series of representative financial scenarios to identify domains where quantum approaches can surpass classical techniques. Furthermore, the paper explores the vulnerabilities quantum computing introduces to digital finance-related applications and blockchain security, including risks to digital signatures, hash functions, and randomness generation, and discusses mitigation strategies through PQC and quantum-resilient alternatives of classical digital finance tools and blockchain architectures. By addressing both quantum blockchain, quantum key distribution (QKD) as well as quantum communication networks, his review presents a more holistic perspective than prior studies, offering actionable insights for researchers, financial practitioners, and policymakers navigating the intersection of quantum computing, blockchain, and secure financial systems.}


\maketitle

\section{Introduction}\label{introduction}
Alongside Artificial Intelligence (AI) and Blockchain, Quantum Computing has emerged as a transformative technology with far-reaching implications across multiple industries. Among the sectors most frequently highlighted for its potential impact is the financial industry, where quantum advances are expected to reshape computational efficiency, risk modeling, and security mechanisms \citep{ORUS2019,herman2022survey,Naik2025}. 

While much of the discourse surrounding quantum computing is framed from a physics or computer science perspective, this paper approaches the subject from a business-oriented standpoint. From this angle, the central question is not merely technological feasibility but rather whether the adoption of quantum computing can generate tangible economic value and profitability.
%Be it that it helps to save costs, generate higher revenues or even open up completely new business models.%
The potential benefits may manifest in various forms, such as lowering operational expenses, enhancing revenue generation, or facilitating the emergence of entirely novel business models. In order to answer this question systematically, we are developing a four-point list of criteria that allows us to derive a decision based on concrete applications in the area of finance. We apply this model to some typical use cases: Quantum Optimization, Quantum Machine Learning, and Quantum Monte-Carlo Simulation. 

Another area of critical importance in the financial sector concerns security challenges, particularly regarding digital banking and blockchain-based technologies, including cryptocurrencies, which are becoming increasingly prevalent \citep{Naik2025}. 
This implies the potential risk that current encryption mechanisms may be rendered obsolete in the near future. Given the highly sensitive nature of banking and insurance data, this concern plays a central role in ongoing discussions, and it will be examined in detail in this study. Blockchain technology has become a foundational framework for decentralized systems, supporting applications such as cryptocurrencies, supply chain management, and secure data sharing. Its security relies on cryptographic primitives—including digital signatures, hash functions, and random number generation—which ensure transaction integrity, authenticity, and immutability\citep{Naik2025}. However, the rise of quantum computing presents significant challenges to these mechanisms. Quantum algorithms, such as Shor’s algorithm \citep{shor1994} for factoring large integers and Grover’s algorithm \cite{GroversSearch} for unstructured search, threaten the security of widely used public-key cryptography like RSA and ECDSA and reduce the effective strength of hash functions \citep{YeniarasFaster,YeniarasImproved}. Given that blockchain systems require long-term security, this poses an urgent need to prepare for the quantum era.

To address these emerging threats, the concept of quantum blockchain has been proposed \citep{QuantumbcXu,QuantumbcLiu,QuantumbcNilesh}. This approach integrates principles from quantum mechanics and quantum information theory to enhance blockchain security and efficiency. Techniques such as quantum key distribution (QKD) \citep{QuantumbcQKDsec}, quantum random number generation (QRNG) \citep{idquantique2018}, quantum-secure communication channels, and quantum digital signatures strengthen authentication, data integrity, and consensus mechanisms. Some designs also leverage quantum entanglement to resist malicious interference, providing resilience against both classical and quantum-capable adversaries \citep{QuantumBlockchains2025}. In addition, quantum-enhanced consensus and mining processes offer potential improvements in speed and energy efficiency.

Despite its promise, implementing quantum blockchain presents substantial challenges. Current quantum hardware faces limitations, including restricted qubit counts, high error rates, and susceptibility to decoherence. Developing quantum software requires specialized knowledge of programming languages like $Q\#$ and $Qiskit$, as well as a deep understanding of quantum mechanics. Furthermore, existing blockchain protocols often need to be modified to accommodate quantum-native consensus mechanisms \citep{QuantumSeureBlockchainQDS} and hybrid classical–quantum frameworks \citep{HybridQuantumbc}. Integrating post-quantum cryptography (PQC) is also critical to secure digital signatures, hash functions, and key management, while balancing performance, storage, and interoperability concerns \cite{Fernandez2024,Quantumresnet}.

We explore the landscape of quantum blockchain research, including prototype implementations, post-quantum cryptographic standards, and the integration of quantum technologies such as QKD for secure communications. By examining both the advantages and limitations of these systems, we highlight the importance of adopting quantum-resistant solutions proactively, ensuring that blockchain infrastructures remain secure, scalable, and reliable in the quantum computing era.

This paper is organised as follows. In chapter (\ref{Intro QC}), we give a very brief and non-technical introduction to quantum computing. This is followed by a chapter (\ref{Relevance}) on the fundamental relevance of quantum computing in finance. The four-stage catalog of business criteria is introduced and discussed here, including a discussion of the associated economic implications. Then chapter (\ref{Use Cases}) applies the proposed decision model to the most prominent foreseeable applications, namely optimization, machine learning, and Monte Carlo simulations.  
Chapter (\ref{Blockchain}) examines the concept of quantum blockchain, focusing on its potential advantages, inherent limitations, and the current landscape of research and practical applicability. While chapter (\ref{Security}) addresses the implications of quantum computing attacks on digital systems and cryptographic primitives within these applications. Within this concept, we analyze possible quantum attack scenarios, review post-quantum algorithms, and discuss quantum-resistant cryptocurrencies as well as architectures for quantum-secure blockchains. The role of Quantum Key Distribution (QKD) in strengthening blockchain security is also highlighted. Finally, chapter (\ref{conclusion}) concludes the paper by providing an outlook on future research directions.

\section{Brief Introduction of Quantum Computing}
\label{Intro QC}

Quantum Computing is an application of Quantum mechanics, a branch of physics that explores the behavior of nature at subatomic scales. Quantum mechanics is intrinsically probabilistic, implying that, in general, statements can only be made with a certain likelihood. The difference between Classical and Quantum Computing stems from the fact that in the Quantum domain, superposition and entanglement play an important role \citep{barletta2023introduction}.
Loosely speaking, before measurement, in general, a quantum system doesn't have a definite value for a property; it exists in a superposed state that encodes all possible outcomes. This can be used in certain quantum algorithms to gain a computable advantage.

Entanglement is another quantum phenomenon in which the joint state of two or more systems cannot be expressed as a product of individual states. As a result, the subsystems exhibit non-classical correlations such that the state of each subsystem cannot be described independently of the others. Even when the entangled systems are spatially separated, a measurement on one subsystem instantaneously determines the outcomes' probability distribution for measurements on the others, consistent with the predictions of quantum mechanics. 

In Quantum Computing, the basic ingredients are the so-called qubits (quantum bits), and all calculations can be described by quantum gates that affect these qubits. In the last thirty years, many quantum algorithms have been developed within this paradigm: Shor's algorithm to factorize large numbers \citep{shor1997polynomial}, Grover's algorithm to perform a search in an unstructured database \citep{GroversSearch}, Quantum Fourier Transformation \citep{Shor2002}, and many more. The most complete collection of quantum algorithms can be found in the Quantum Algorithm Zoo: \url{https://quantumalgorithmzoo.org/}

For about 10 years now, it has also been possible not only to investigate quantum algorithms theoretically or to simulate them classically on a small scale, but also to execute them on real quantum computers. Despite enormous technological progress in recent years, these still only consist of a moderate number of qubits. Furthermore, the actual calculations must not take too long, otherwise several errors on the qubit level will destroy the results. People call this NISQ era; NISQ stands for noisy intermediate-scale quantum computing \citep{NISQQuantum}. 

With quantum algorithms, a distinction is therefore regularly made as to whether they are (also) suitable for the current NISQ computers or whether they require a (large) number of error-free qubits that can be processed for a sufficiently long time. Popular NISQ algorithms are variational approaches such as QAOA \citep{farhi2014quantum} or approaches like Quantum Neural Networks or Quantum Support Vector Machines, which play a dominant role in the domain of Quantum machine learning \citep{schuld2018supervised}.

In Quantum  Technology research and development, the ultimate goal, however, is to have (hundred) thousand error-free qubits available and to perform calculations as long as necessary. To achieve this, big companies like IBM, Google, and Microsoft, but also quantum start-ups with different approaches, are in a fierce technological competition. Now, in 2025, many people in the field think that in 2030, we may do calculations with about 1000 error-free qubits.      

In addition to the technological efforts to build large quantum computers, intensive research has been carried out for several years into the specific applications for which quantum computing is suitable. One area that is always mentioned here is the financial sector. 

\section{Economic Relevance of Quantum Computing in Finance} \label{Relevance}
\subsection{Efficient solution of finance problems}
Finance, as a strongly data-driven business, is full of complicated and time-consuming calculation problems. These range from the valuation of derivatives and risks to the question of optimal portfolio allocation and scenario calculations in the insurance sector. Companies that solve these calculations the fastest, most accurately, or most efficiently have long held a comparative advantage in the market. It therefore makes sense to consider finance as a potential area of application for quantum computing.

To assess the existing advantages of quantum computing more precisely, it is, however, advisable to carry out the following four-stage analysis:

\subparagraph {Stage 1}Which computational problem is currently solved unsatisfactorily, too inaccurately, too slowly, too inefficiently?
\subparagraph {Stage 2} Is there a quantum algorithm (or more generally, a quantum method) that is applicable?
\subparagraph {Stage 3} Can a computational advantage (accuracy, speed, efficiency, etc) be achieved with this quantum algorithm/method?
\subparagraph {Stage 4} Does this result in an economic business case, i.e., can an economic advantage be achieved in the overall consideration of total costs and benefits compared to the classical (i.e., non-quantum) alternative?

A real quantum advantage for users in companies will only arise when there is a positive answer to all of these questions. This statement is valid for all industries and can be applied to finance as well.

If we look at the current discussions about quantum benefits in the financial sector (see e.g. \citep{bis2024} for a comprehensive compilation of the status quo), you can see that regularly only questions 2 and 3 are considered in more detail. 
In other words, it is not sufficiently considered whether there is currently a real problem at all and whether a quantum approach pays off in the overall view. Almost everything revolves around the question of quantum algorithms and their generic behavior in terms of runtime, accuracy, etc. 

In detail, three different constellations can be distinguished for the algorithmic behavior:
{\bf Algorithms with a theoretical exponential speedup,} e.g., Shor's algorithm. These solve certain problems in polynomial time instead of exponential time. Whereas Shor's algorithm is primarily designed to factorise large numbers into primes, part of the algorithm (e.g., Quantum Fourier Transformation) can be applied to other and maybe more relevant questions. 
{\bf Algorithms that only allow a quadratic speedup} (“10 sec instead of 100 sec”), see e.g., Grover's algorithm \citep{GroversSearch}. Although there is a whole series of economically interesting use cases for these algorithms, it is disputed whether the benefit of a quadratic speedup is sufficient to justify the effort/costs in the overall view (see stage 4 above).
{\bf Many other quantum algorithms,} e.g., in the field of optimization or quantum machine learning, for which there is no theoretical proof of a speedup or an accuracy advantage. Many research papers investigate this question empirically and also use data or problems from the field of finance.
    
The typical application problems discussed in the section (\ref{Use Cases}) are mainly applications of the last-mentioned algorithms, since those seem to fit best for tasks in the financial domain. This means, however, that in most cases, the question if a quantum advantage exists, how large it may be, and whether an application is really worthwhile in the overall context is an open question

\subsection{Economic advantages in a broader sense}

As soon as an economic quantum advantage is achieved at the individual company level for distinct use cases (i.e., the questions in stages 1-4 above are answered positively), these advantages can be transferred to broader systemic and macroeconomic dimensions (possibly with gradations). There are hardly any economic studies in this direction in the literature to date. One of the few examples is the work by Bova et al \citep{Bova2023}, in which a Cournot duopoly is modeled, where a quantum computing company competes against a classical computing company. 

\noindent
However, broader economic consequences arise in particular for the following topics:

\subparagraph{First-Mover Advantage} Institutions that invest in quantum capabilities early may benefit from a durable competitive advantage \citep{Bova2023}. Quantum-native trading algorithms, optimization heuristics, and fraud detection models can deliver superior performance \citep{Egger_2020}. These advantages are compounded by the difficulty of replicating such capabilities due to the steep learning curve and capital requirements.
 
\subparagraph{Market Efficiency Gains} Traditional markets operate under the assumption that prices reflect available information (Efficient Market Hypothesis). However, in practice, processing delays and computational bottlenecks allow short-lived arbitrage opportunities. Quantum-enhanced analytics can drastically shorten information processing times, thereby reducing latency and making markets more information-efficient \citep{Zhuang_2022}.

\subparagraph{Innovation Diffusion} As quantum solutions become commoditized and accessible via quantum cloud services, smaller financial institutions and emerging markets may gain access to advanced computational capabilities. This democratization can help level the playing field, encouraging innovation between geographies and organizational sizes. Market-wide adoption could catalyze innovative financial products, better risk models, and more appropriate regulatory frameworks \citep{Diffusion}.

However, quantum development requires highly specialized expertise in physics, computer science, and financial engineering. The scarcity of quantum talent and the high costs of R\&D investments present substantial barriers to entry. Consequently, access to quantum expertise is likely to be concentrated among top-tier firms and research institutions in the near term. Furthermore, legacy banking systems and risk management frameworks are not readily compatible with quantum computing paradigms. The interface of quantum tools with classical workflows requires robust middleware, hybrid algorithms, and modifications in data architecture. Developing such integrations involves technical complexity and operational risk.

\section{Quantum Optimization, Quantum Machine Learning, Quantum Monte Carlo Simulation}\label{Use Cases}
In the finance domain, many people consider the following use cases as highly relevant as potent applications of quantum computing: Optimization, Quantum Machine Learning, and Monte Carlo Simulation, cf. \citep{Naik2025}. In each of the following subparagraphs, we adopt the scheme from above and answer the questions in stages 1 to 4 for these applications.

\subsection{ Quantum Optimization}

\subparagraph {Stage 1} Optimization problems, especially combinatorial optimization problems, can be found in various areas of finance: from portfolio optimization and securities settlement to accounting problems. While a faster calculation of the securities to be matched for settlement is desirable in securities settlement, the advantages of a faster portfolio optimization are not so obvious. Furthermore, it is debatable whether a perfect global solution is always necessary in practice or if classical optimization heuristics available are sufficient. 

\subparagraph {Stage 2}The quantum algorithms that are suitable here are mainly Grover approaches with a quadratic speedup, as well as empirical variational approaches, especially the QAOA algorithm. In a popular list of QUBO (Quadratic unconstrained binary optimization problem) problems (\url{https://blog.xa0.de/post/List-of-QUBO-formulations/}), many problems from the financial industry appear, see also \citep{martini2025,cohen2020portfolio2,mattesi2023financial,accounting}.

\subparagraph {Stage 3}Although Grover algorithms offer a theoretical speedup, they require a large number of error-free qubits for their application, meaning that practical use will not be possible in the next couple of years.  However, it should be noted that classical algorithms have an exponential run-time for this problem class, and the Grover algorithm only provides a quadratic speedup. Thus, the runtime remains exponential and is not reduced, as it would be desirable, to a polynomial runtime. Algorithms that are already executable on quantum hardware available today are mostly from the category of variational algorithms, like QAOA. Here, there is the hope that, at least for certain problem instances with determined properties (e.g., symmetries, etc.), runtime advantages over the classic counterparts can be achieved. While some results seem to point in this direction, the biggest problem with the current research is that the number of qubits considered is still too small for a conclusive answer \citep{Dehn2025}.

 \subparagraph {Stage 4} Particularly in view of the fact that a clear quantum advantage at the algorithmic level has not yet been proven beyond doubt, the question of whether there is a business advantage of applications in the overall view is currently open.

\subsection{Quantum Machine Learning}

    \subparagraph {Stage 1} Machine learning algorithms can be found in many different positions in financial companies. Starting with the rating and scoring of credit or insurance customers, through fraud detection and customer segmentation, to autonomous trading systems. In principle, all three main areas of machine learning, supervised learning, unsupervised learning and reinforcement learning, are covered. A higher accuracy or discriminatory power of the established procedures is desirable in most cases and translates into higher profits or lower costs.

  \subparagraph {Stage 2}In the domain of quantum machine learning at present, two main approaches dominate \citep{schuld2018supervised}: Quantum Support Machines and Quantum Neural Nets. Both are strongly inspired by their classical counterparts in a way that certain elements of the algorithm is substituted with a quantum element. E.g., in Quantum Neural Networks, the "neurons" are quantum gates with parameters to be trained.  
        
  \subparagraph {Stage 3}  Here, a quantum advantage would not necessarily mean a speed-up of the calculation (training or inference) with the same amount of data, but would mean achieving better generalization models. This could lead to shorter training times or the usage of less data, cf. e.g. \citep{Hellstern2021}.
    The problem of determining which quantum machine learning model is successful for which kind of data is highly investigated and partially unsolved. It seems that the "no free lunch theorem" \citep{Nofreelunch} known from classical machine learning persists in the quantum domain. From a practical point of view, this means that any advantage may be highly dependent on the data.
    
    \subparagraph {Stage 4} As soon as a quantum advantage is found, business benefits can be examined in more detail. To answer this, the cost of achieving the higher accuracy etc, will play a crucial role. Although more accurate rating systems lead to lower default risks and operational risks can be reduced with better fraud detection procedures, it is not rational to spend a vast amount of money if the gain is only marginal.

\subsection{Quantum Monte Carlo Simulation}

\subparagraph {Stage 1} Monte Carlo simulations are used in financial institutions to evaluate complex banking and insurance products and also serve as a flexible method for calculating risks. However, in order to achieve the most accurate results possible, a large number of simulations are necessary, which is reflected in regularly long calculation times. In this respect, it would be highly desirable if it were possible to reduce the number of simulations required to achieve a certain level of accuracy. For example, in the trading area of banks, every reduction in time means a comparative advantage.

\subparagraph {Stage 2}In 2018, Rebentrost et al. \citep{Rebentrost_2018} proposed an algorithm that demonstrated Monte Carlo simulations for finance using a quantum computer. This consists of a sequence of several calculation steps as soon as the problem specifications (e.g., underlying distribution) are encoded as quantum states and is based on earlier work of \citep{Brassard_2000, Montanaro_2015}. In the meantime, this approach has been improved and made more efficient \citep{Egger_2020_1}.

\subparagraph {Stage 3}The construction of the quantum Monte Carlo approach theoretically leads to a quadratic speedup, i.e., if $N$ simulation runs are classically necessary for a given error, then $\sqrt{N}$ simulation runs of the quantum algorithm are sufficient. When applying the algorithm to real-world problems, cf. e.g. \citep{Stamatopoulos2020, Chakrabarti2021}, it becomes evident that it is necessary to have at least several thousand error-free qubits to perform many calculation steps. This is presently far beyond the technical possibilities and may not be possible before the next decade.

\subparagraph {Stage 4}  As stated above, to calculate the prices of exotic derivatives much faster than now will offer an advantage for the big investment banks. More accurate risk calculations will be beneficial for all institutions. As both topics involve complex and data-intensive calculation processes at the institutes, it will be necessary to examine very closely where it makes sense to replace classical processes with their quantum counterparts. Similarly to the optimization and machine learning case, any advantage must offset potentially higher implementation and production costs of quantum methods.

\section{Quantum Blockchain}\label{Blockchain}

Quantum blockchain is a decentralized, encrypted ledger designed to safeguard blockchain networks from potential threats posed by quantum computers \citep{QuantumBlockchains2025,QuantumbcLiu}. Leveraging principles from quantum computing and quantum information theory, it enhances security through quantum cryptographic methods such as quantum key distribution (QKD) \citep{QuantumbcQKDsec}, quantum random number (QRNG) generation \citep{idquantique2018}, and quantum-secure communication channels \citep{Quantumresnet}. Some proposed models even employ quantum entanglement---where particles remain correlated despite never interacting---to strengthen consensus mechanisms \citep{QuantumConsensus} and prevent malicious interference.

Like traditional blockchains, quantum blockchains maintain decentralization but offer superior resilience and efficiency. Techniques such as quantum digital signatures \citep{quantumdigitalsig2,quantumdigitalsignatures,QuantumSeureBlockchainQDS} ensure authentication and data integrity, protecting against attacks from quantum-capable adversaries. Future implementations may also integrate algorithms like Grover’s for more efficient mining \citep{QBT2025}, though this raises fairness concerns until quantum hardware becomes more accessible.

\subsection{Advantages of Quantum Blockchain}
The advantages and potential of quantum blockchain span several critical domains. Although fully functional quantum computers have not yet been realized, integrating quantum principles into blockchain architectures presents promising avenues for transformation. Security can be substantially enhanced through quantum key distribution (QKD) \citep{QKDbc1,QuantumSeureBlockchainQDS,QKDbc3,QuantumbcQKDsec} and quantum cryptography \citep{quantumdigitalsig2, idquantique2018} which protect transactions from decryption by quantum-enabled adversaries \citep{Fernandez2024, Quantumresnet}. Consensus processes may also accelerate, as quantum superposition and entanglement enable faster transaction validation \citep{QuantumConsensus}. Efficiency gains are anticipated as optimized quantum algorithms could reduce transaction times and lower mining costs.

Furthermore, quantum-powered analytics and optimization have the potential to revolutionize sectors such as finance, healthcare, and logistics by facilitating more sophisticated data processing. Quantum mining \citep{QBT2025}, particularly through proof-of-quantum-work \citep{Amin2025} protocols, could surpass classical mining in speed and energy efficiency—some estimates suggest up to 1,000-fold reductions in energy consumption—though such advantages may introduce centralization risks if control is concentrated. Overall, preparing blockchain systems today for a quantum-driven cybersecurity environment ensures resilience in the future.

\subsection{Limitations and Challenges of Quantum Blockchain}
Despite its promise, quantum blockchain faces significant challenges that must be addressed before practical deployment. Hardware remains a major constraint: current quantum computers require large numbers of stable qubits with extremely low error rates, a technological and financial hurdle \citep{Cai2022, Naik2025}. Additionally, operating within the NISQ era, systems are limited by decoherence and noise, which restrict computational depth and coherence time \citep{NISQQuantum, NISQ2}.

Programming complexity is another barrier. Developing quantum software requires expertise in specialized languages such as Q\# and Qiskit, along with a strong understanding of quantum mechanics—skills that are still relatively scarce, although expanding open-source resources may gradually mitigate this limitation. Algorithmic compatibility also poses challenges: many classical blockchain protocols are not readily adapted to quantum systems, necessitating the design of quantum-native consensus mechanisms, which simultaneously represents an opportunity for innovation \citep{Wen2022}.

Reliability concerns persist due to noise and decoherence; however, progress in quantum error correction and fault-tolerant designs, including concatenated error-suppressing codes, offers a pathway toward stability \citep{Quantinuum2025, Cai2022}. Integrating classical and quantum systems introduces further complexity for security and interoperability, though hybrid \citep{HybridQuantumbc} approaches may ultimately leverage the advantages of both paradigms.

Institutionally, the absence of standardized frameworks, regulatory clarity, and interoperable protocols slows adoption but provides a chance for policymakers and industry leaders to influence governance from the outset. Access remains limited as large-scale quantum infrastructure is currently restricted to well-funded organizations, though cloud-based quantum services could lower entry barriers. From a cybersecurity perspective, quantum technologies both strengthen blockchain security and introduce new vulnerabilities, highlighting the need for post-quantum cryptography and adaptive protocols \citep{Quantumresnet, Fernandez2024}.

\subsection{Current Research and Prototypes of Quantum Blockchain}
At present, no fully operational quantum blockchain exists; most initiatives  remain theoretical or focus on quantum-resistant cryptographic solutions \citep{Naik2025, Fernandez2024, Yang2024}. While these ledgers provide security against classical threats, their resilience against future quantum attacks remains uncertain \citep{Wang2022}. Research in quantum-resistant encryption continues to advance, alongside efforts to integrate quantum computing into financial systems and address regulatory and interoperability concerns \citep{Amin2025, Quantumresnet}.

Several prototype quantum blockchain models \citep{QuantumbcLiu,QuantumbcNilesh,QuantumbcXu,QuantumbcQKDsec,QuantumBlockchains2025,HybridQuantumbc,QuantumSeureBlockchainQDS} have been proposed in both academia and industry. In \citep{Amin2025} a quantum proof-of-work protocol, highlighting potential efficiency and security benefits is introduced. In industry, Quantum Blockchains Inc. has developed the Quantum Secured Blockchain (QSB) \citep{QuantumBlockchains2025}, while Quantum Blockchain Technologies Plc is exploring quantum-enhanced blockchain mining solutions \citep{QBT2025}. These prototypes aim to test the practical implications of quantum technologies on consensus mechanisms, transaction throughput, and cryptographic security, even though fully functional systems remain a goal for the future.

If challenges in hardware, algorithmic design, error correction, and system integration can be addressed, quantum blockchain has the potential to significantly enhance scalability, security, and efficiency in financial networks and beyond \citep{Fernandez2024, Yang2024}.

\section{Quantum Security: Attacks and Mitigation Strategies on Blockchain Systems}\label{Security}

Blockchain technology relies on cryptographic primitives—digital signatures, hash functions, and random number generation—to preserve decentralization, authenticity, and immutability \citep{Naik2025}. The emergence of quantum computing poses a fundamental threat to these foundations, as it leverages principles such as superposition and entanglement to perform certain classes of computations exponentially faster than classical machines. Among the most notable breakthroughs, Shor's algorithm \citep{shor1994,shor1997polynomial,Shor2002,bernstein2009} can efficiently factor large integers and solve discrete logarithms, thereby undermining the public-key cryptography that secures most blockchain transactions. In parallel, Grover's algorithm \citep{GroversSearch} accelerates unstructured search problems, effectively reducing the strength of symmetric encryption and hash functions by a square root factor. Together, these developments highlight how both transaction validation and data integrity mechanisms could be severely weakened in a post-quantum era \citep{Naik2025}.

Although current quantum computers remain far from the scale necessary to break blockchain cryptography in practice, the pace of progress suggests that this risk cannot be dismissed as purely theoretical. Moreover, blockchains are characterized by their long-term security requirements: transactions recorded today must remain verifiable decades into the future. This temporal asymmetry underscores the urgency of transitioning toward quantum-resistant solutions, with lattice-based, hash-based, and code-based cryptographic schemes emerging as the most promising candidates. The sustainability of blockchain systems will therefore depend on a timely and technically rigorous adoption of these post-quantum defenses.

\subsection{Shor's Algorithm Attacks on Public Key Cryptography and Digital Signatures}
Digital signatures are based on public key cryptography and they are used to authenticate transactions and validate participants in blockchain networks. Common signature schemes such as ECDSA, RSA, and EdDSA rely on the hardness of the Elliptic Curve Discrete Logarithm Problem (ECDLP) and the Integer Factorization Problem (IFP) \citep{Naik2025}. Shor's quantum factoring algorithm can efficiently solve these problems, enabling quantum adversaries to:
\begin{itemize}[label=\tiny$\bullet$]
    \item Derive private keys from public keys.
    \item Forge signatures for transactions or smart contract executions.
    \item Impersonate validators in Proof of Stake (PoS) systems.
    \item Manipulate smart contracts dependent on signature-based permissions.
\end{itemize}

In PoS (Proof of Stake) systems, the reliance on digital signatures for validator selection and block signing makes them particularly susceptible to quantum attacks. A quantum adversary could compromise the consensus mechanism by impersonating legitimate validators, undermining the network's integrity.

\subparagraph{Long-term risk} Even before large-scale quantum computers are available, adversaries could record blockchain data and smart contract interactions today. Once quantum computing capabilities advance, these recorded interactions could be decrypted or signatures forged, leading to a "harvest now, exploit later" scenario that threatens both financial assets and contract integrity.

\subparagraph{Mitigation} Transitioning to post-quantum signature schemes, such as lattice-based or hash-based algorithms, is essential to safeguard against quantum attacks. Implementing these schemes proactively can ensure the continued security of blockchain systems \citep{Naik2025}.

\subsection{Hash Functions, Grover's Algorithm, Proof of Work, and Hash-Based Smart Contract Risks}
Hash functions are integral to blockchain operations, linking blocks, securing Merkle trees, and underpinning Proof of Work (PoW) mechanisms. Grover's algorithm provides a quadratic speedup for brute-force search problems, reducing the complexity of finding preimages from $2^n$ to $2^{n/2}$.

For SHA-256, this implies a reduction in effective security from 256 bits to 128 bits. While still substantial, this reduction has several implications:

\begin{itemize}[label=\tiny$\bullet$]
    \item Quantum miners could find valid hashes more efficiently, potentially centralizing mining power.
    \item Hash-based smart contracts, such as those involving time-locks or conditional payments, may experience reduced effective security, shortening their reliability.
    \item Long-term exposure exists for contracts or commitments recorded today, which could be exploited by future quantum attacks.
\end{itemize}

\subparagraph{Mitigation}  Employing longer-output hash functions, such as SHA-512, or transitioning to quantum-resistant hash constructions can help maintain security margins and ensure the durability of hash-based systems.

\subsection{Random Number Generation (RNG) and Smart Contract Vulnerabilities}
Secure randomness is critical for nonce selection, validator lotteries, and smart contract operations. Weak or biased RNG has historically led to vulnerabilities, such as the 2013 Android Bitcoin wallet incident, where predictable randomness resulted in private key leakage.

Quantum adversaries could exploit RNG weaknesses by:

\begin{itemize}[label=\tiny$\bullet$]
    \item Accelerating nonce or random selection prediction in PoW or signature schemes using Grover's algorithm.
    \item Biasing validator lotteries in PoS networks.
    \item Manipulating random outcomes in smart contracts, such as lotteries or conditional functions.
\end{itemize}

\subparagraph{Mitigation}  Robust algorithmic RNGs should be combined with quantum random number generators (QRNGs) \citep{idquantique2018} that leverage quantum physical phenomena to produce truly unpredictable outputs, ensuring fairness and the integrity of smart contracts.

\subsection{Post-Quantum Cryptography (PQC)}
Post-Quantum Cryptography (PQC) \citep{NISTPQCProcess, Naik2025} seeks to develop cryptographic algorithms that can withstand potential attacks from quantum computers. While quantum systems capable of compromising widely used schemes such as RSA-2048 or elliptic curve cryptography (ECC) have not yet been realized, experts emphasize the importance of proactive preparation. This urgency stems from several factors: blockchain data is susceptible to a “store-now, decrypt-later” threat model, the migration of cryptographic protocols requires consensus across entire networks and extensive software updates, and ensuring backward compatibility with legacy transactions remains a critical challenge.

\subsubsection*{Algorithm Families}

The field of post-quantum cryptography (PQC) has developed rapidly in response to the emerging threat posed by quantum computers to conventional public-key cryptographic schemes. To address this challenge, the National Institute of Standards and Technology (NIST) launched its PQC standardization process in 2016 \citep{NISTPQCProcess}, which remains ongoing and has become the authoritative global benchmark for evaluating candidate algorithms. The main categories of PQC algorithm families can be summarized as follows:

\begin{itemize}[label=\tiny$\bullet$]
    \item \textbf{Lattice-based}: Kyber (KEM), Dilithium (signature), Falcon, NTRU, FrodoKEM --- efficient, versatile, and among the most widely adopted candidates.
    \item \textbf{Hash-based}: XMSS, SPHINCS+ --- provably secure with strong theoretical guarantees, though often associated with larger key and signature sizes.
    \item \textbf{Code-based}: Classic McEliece, BIKE, HQC --- long-established robustness, but hindered by very large public key sizes.
    \item \textbf{Multivariate polynomial-based}: Rainbow, GeMSS, Picnic --- enable fast verification, though their security resilience has varied over time, with several notable cryptanalytic breaks.
    \item \textbf{Symmetric-based}: FAEST --- relies on well-understood symmetric primitives, providing a conservative and security-grounded approach.
    \item \textbf{Supersingular isogeny-based}: SIKE --- built on hard problems in elliptic curve isogeny graphs, though recent attacks have significantly undermined its security.
\end{itemize}

In 2022, NIST announced the first standardized PQC algorithms: Kyber for key encapsulation and Dilithium for digital signatures, while also advancing Falcon and SPHINCS+ as additional signature schemes. At the same time, research into other algorithm families continues, ensuring a diversified portfolio of cryptographic tools to achieve resilience against future quantum adversaries.

\begin{table}[htp]
    \setlength{\tabcolsep}{6pt}
    \centering    
    \caption{Set of Post-Quantum Cryptography Standards by NIST's Announcement (2022-2025)}
     \label{tab:1}
    \label{tab:NISTannounce}
    \begin{tabular}{|l|l|}
        \hline
        \textbf{Key Establishment Standards} & \textbf{Digital Signature Algorithm Standards} \\ 
        \hline
        CRYSTALS-KYBER & CRYSTALS-Dilithium \\ 
        HQC (backup)   & FALCON \\ 
                       & SPHINCS+ \\ 
        \hline
    \end{tabular}
\end{table}
\FloatBarrier

NIST issued a request for proposals for additional post-quantum signature schemes \cite{NISTAdditionalPQCSignatures,Naik2025} to enhance their initial selection. By 2023, they had received 50 submissions, of which 40 were deemed suitable as Round-1 candidates for potential future standardization. Currently, as of 2025, Round-2  of those digital signature algorithms are under evaluation.

\subsection{Quantum-Resistant Blockchains}
A quantum-resistant blockchain replaces cryptographic primitives that are vulnerable to quantum attacks with post-quantum cryptography (PQC) alternatives, thereby safeguarding assets and transactional integrity against emerging quantum computing threats \citep{Naik2025}. Deployment strategies for implementing PQC in blockchain systems vary: some networks may employ hard forks to introduce PQC-based transaction formats, ensuring that all future transactions utilize quantum-secure signatures, while others adopt hybrid models that support both classical and PQC signatures during the migration period, allowing a gradual transition and reducing disruption to existing users. Proof-of-Burn mechanisms can also be employed to securely transition assets between classical and quantum-resistant frameworks, mitigating the risk of double-spending or transaction conflicts.

The integration of PQC into blockchain systems presents several technical challenges. Transaction sizes can increase dramatically, as PQC signatures are typically five to fifty times larger than classical ECDSA signatures, which can lead to slower propagation and higher bandwidth requirements. Computational costs for verification and block propagation are also higher, potentially impacting throughput and network efficiency. Moreover, the larger key and signature sizes contribute to ledger growth, exacerbating storage demands and necessitating protocol-level adjustments or layer-2 scaling solutions. 

The need to switch to quantum-resistant cryptocurrencies is particularly urgent because quantum computers capable of breaking widely used elliptic-curve and RSA-based signatures could render current blockchains insecure, exposing billions of dollars in assets to theft or forgery. Even if large-scale quantum computers are still years away, the “store now, decrypt later” threat implies that adversaries can capture encrypted transactions today and compromise them in the future once quantum capabilities become available. Failure to adopt PQC proactively could therefore result in irreversible financial losses, undermine trust in decentralized networks, and slow the evolution of blockchain-based applications, highlighting the critical importance of immediate research, testing, and deployment of quantum-resistant cryptographic protocols. In Table 2 some recent examples of existing quantum-resistant cryptocurrencies \citep{Naik2025, AlgorandPostQuantum,KomodoDilithiumIntegration,CellframeQuantumSafe,QANplatformQuantumResistant} are listed.

\begin{table}[htp]
    \setlength{\tabcolsep}{6pt}
    \centering    
    \caption{Cryptocurrencies with Quantum-Resistant Digital Signature Algorithms}
    \label{tab:2}
    \begin{tabular}{|l|l|}
        \hline
        \textbf{Cryptocurrency} & \textbf{Quantum-Resistant Signatures} \\ 
        \hline
        enQlave & XMSS \\ 
        Hcash & BLISS \\ 
        IOTA & WOTS (Winternitz one-time signature scheme) \\ 
        Mochimo & WOTS (Winternitz one-time signature scheme) \\ 
        QAN & XMSSMT (hybrid) \\ 
        QRL & XMSS \\ 
        Quantum Resistant Coin (QRC) & XMSS \\ 
        Nexus & FALCON \\ 
        Algorand (ALGO) & FALCON \\ 
        Cellframe (CELL) & Dilithium \\ 
        Komodo (KMD) & Dilithium \\ 
        QaN Platform (QANP) & Dilithium \\ 
        \hline
    \end{tabular}
\end{table}
\FloatBarrier

\subsection{Quantum-Secure Blockchain Architecture}
It is crucial to differentiate between quantum-resistant and quantum-secure blockchains. A quantum-resistant blockchain is primarily concerned with replacing cryptographic components that are vulnerable to quantum attacks with post-quantum cryptographic (PQC) schemes such as XMSS, FALCON, or Dilithium. This substitution strengthens the security of digital signatures, hash functions, and occasionally randomness sources, thereby improving resilience against adversaries equipped with quantum computational capabilities. By contrast, a quantum-secure blockchain \citep{Naik2025,QuantumSeureBlockchainQDS} takes a holistic approach, aiming to neutralize quantum threats across all structural layers of the system. This broader model includes adapting consensus mechanisms such as Proof-of-Work (PoW) and Proof-of-Stake (PoS) to prevent advantages from quantum acceleration, deploying post-quantum secure communication protocols to safeguard inter-node exchanges from quantum-assisted man-in-the-middle attacks, and employing Hardware Security Modules (HSMs) that are compatible with PQC for more reliable key management \citep{quantumsecurebc1,Quantumresnet,nist_nccoe_migration_pqc_2025}.

In addition, the use of Quantum Random Number Generators (QRNGs) \citep{idquantique2018}for validator selection introduces genuinely unbiased randomness, which is essential for protecting consensus processes against manipulation by either classical or quantum adversaries. A further promising development is the application of Quantum Key Distribution (QKD) to blockchain networking \citep{QKDbc1,QuantumSeureBlockchainQDS,QKDbc3}. Leveraging the principles of quantum mechanics, QKD enables theoretically secure key exchange that remains uncompromised even by adversaries with unbounded computational resources. Within blockchain ecosystems, QKD could be applied to secure validator coordination, protect peer-to-peer communication channels, or reinforce protocols such as Transport Layer Security (TLS), which serve as the foundation of blockchain networking stacks \citep{QuantumbcQKDsec,quantumsecQKDTLS1,quantumsecQKDTLS2}. Some researchers have also suggested hybrid frameworks in which QKD generates session keys, while classical cryptographic methods continue to handle the bulk of data transmission for scalability and efficiency.

The urgency of these advancements is underscored by the disruptive potential of quantum computing, which threatens to render current signature schemes such as RSA and ECDSA insecure and to weaken the robustness of widely used hash functions. Transitioning to PQC solutions \citep{nist_nccoe_migration_pqc_2025,moody2024transition} must therefore be approached with careful consideration of interoperability, performance overheads, and storage requirements. Early integration of PQC standards, quantum-secure networking protocols—including QKD—and robust sources of randomness can help ensure that blockchain infrastructures retain their security guarantees in the quantum era. Although migration will involve significant financial and technical costs, the risks of inaction—such as mass asset compromise or the erosion of trust in decentralized systems—are far more severe.

\section{Conclusion }\label{conclusion}

In this paper, we examined the emerging role of quantum computing in the financial sector, emphasizing its potential advantages over classical methods from a business perspective. We proposed a four-step criteria catalog to evaluate whether a particular application of quantum computing could generate tangible benefits, such as cost reduction, improved revenue, or the creation of entirely new financial opportunities. The paper analyzed key applications, including portfolio optimization, risk management, derivatives pricing, and machine learning for predictive analytics, highlighting the theoretical speedups and efficiency gains offered by quantum algorithms. Importantly, our review goes beyond typical analyses by also considering blockchain technology and post-quantum cryptography (PQC), providing a more comprehensive perspective on the convergence of quantum computing, cybersecurity, and finance. We explored quantum-secure architectures, quantum-resistant cryptography, and the challenges posed by algorithms such as Shor’s and Grover’s. Our review emphasizes that while quantum technologies promise transformative improvements in computation and security, practical deployment remains constrained by current hardware limitations, noise, and error correction challenges. Nonetheless, proactive adoption of PQC, integration of quantum-random number generators, and hybrid classical–quantum approaches can prepare financial institutions for the eventual arrival of large-scale quantum machines. Overall, the study demonstrates that quantum computing has significant potential to reshape financial processes, but careful assessment, strategic investment, and ongoing research are necessary to convert theoretical advantages into real-world gains.
\subparagraph {Future Research Directions} Building on the insights from this review, several areas warrant further investigation to translate the potential of quantum computing into practical applications in finance and secure blockchain systems. Research should explore the development of quantum algorithms compatible with current NISQ devices, targeting portfolio optimization, risk modeling, and derivatives pricing, with a focus on error resilience and computational efficiency. Quantum-enhanced machine learning, including hybrid classical–quantum models, could advance predictive analytics, fraud detection, and algorithmic trading.

On the security front, the practical integration of post-quantum cryptography (PQC) into blockchain networks requires further study, addressing performance optimization, ledger storage management, and hybrid migration strategies. Fully quantum-resilient blockchain architectures, incorporating secure consensus protocols, quantum-random-number-based validator selection, and quantum-safe communication channels, represent another critical area for research.

Additionally, cost-benefit analyses examining the trade-offs between quantum adoption, implementation costs, and measurable financial gains will help guide strategic decision-making for institutions. Finally, the development of regulatory and standardization frameworks, including industry-wide interoperability and governance of quantum-secure protocols, will be essential for ensuring a robust, secure, and harmonized financial ecosystem in the quantum era. Addressing these topics will bridge the gap between theoretical advancements and practical deployment, enabling institutions to harness quantum technologies effectively while maintaining security against emerging quantum threats.
\backmatter

%\bmhead{Supplementary information}

\bmhead{Acknowledgements}
G.H. thanks the organizers of the "Practical Applied Research Conference (PARC)" at the Dublin Business School for the opportunity to present this work.

\section*{Declarations}

\begin{itemize}
\item Funding:
This work is funded by the German Federal Ministry of Research, Technology and Space
within the funding program
‘Application-oriented quantum computing’ under Contract No. 13N17159.

\item Conflict of interest/Competing interests: 
The authors have no relevant financial or non-financial interests to disclose.

\item Ethics approval and consent to participate:
Not applicable. 

\item Consent for publication:
Not applicable.

\item Data availability:
Not applicable.

\item Materials availability:
Not applicable.

\item Code availability:
Not applicable.

\item Author contribution:
The authors contributed equally to the paper. All authors read and approved the final manuscript.
\end{itemize}

\bibliographystyle{agsm} % Harvard  bstyle
\bibliography{Main} 

%\end{thebibliography}
\end{document}